\documentclass[conference,10pt,letterpaper]{IEEEtran}
\IEEEoverridecommandlockouts
\usepackage[style=ieee,citestyle=numeric-comp,sorting=nyt,url=false]{biblatex}
\usepackage{amsmath,amssymb,amsfonts}
\usepackage{graphicx}
\usepackage{textcomp}
\usepackage{tabularx}
\usepackage[hidelinks]{hyperref}
\usepackage[T1]{fontenc}
\usepackage{minted}
\usepackage[defaultmono]{droidsansmono}
\usepackage{capt-of}

\begin{document}

\title{SD3: The First Operationally-Proven, Fully Heterogeneous-Accelerated Global Kilometer-Scale Model with Inline Physics Architecture}

\author{
\IEEEauthorblockN{Yuan Liang\IEEEauthorrefmark{1}}
\IEEEauthorblockA{TianJi Weather Science \\
and Technology Company \\
Beijing, China \\
yuan\_liang@tjweather.com}
\and
\IEEEauthorblockN{Xiaomin Hu}
\IEEEauthorblockA{TianJi Weather Science \\
and Technology Company \\
Beijing, China \\
xiaomin\_hu@tjweather.com}
\and
\IEEEauthorblockN{Xi Chen\IEEEauthorrefmark{1}}
\IEEEauthorblockA{State Key Laboratory of Earth System Numerical \\
Modeling and Application, \\
Institute of Atmospheric Physics, \\
Chinese Academy of Sciences \\
Beijing, China \\
chenxi@lasg.iap.ac.cn}
\and
\IEEEauthorblockN{Jingsheng Zhu}
\IEEEauthorblockA{TianJi Weather Science \\
and Technology Company \\
Beijing, China \\
jingsheng\_zhu@tjweather.com}
\and
\IEEEauthorblockN{Li Yang}
\IEEEauthorblockA{TianJi Weather Science \\
and Technology Company \\
Beijing, China \\
li\_yang@tjweather.com}
\and
\IEEEauthorblockN{Leisheng Li}
\IEEEauthorblockA{Institute of Software, \\
Chinese Academy of Sciences \\
Beijing, China \\
leisheng@iscas.ac.cn}
\and
\IEEEauthorblockN{Qian Song}
\IEEEauthorblockA{TianJi Weather Science \\
and Technology Company \\
Beijing, China \\
qian\_song@tjweather.com}
\and
\IEEEauthorblockN{Xu Wang}
\IEEEauthorblockA{TianJi Weather Science \\
and Technology Company \\
Beijing, China \\
xu\_wang@tjweather.com}
\and
\IEEEauthorblockN{Hao Wang}
\IEEEauthorblockA{TianJi Weather Science \\
and Technology Company \\
Beijing, China \\
hao\_wang@tjweather.com}
\and
\IEEEauthorblockN{Zhenzhen Wang}
\IEEEauthorblockA{TianJi Weather Science \\
and Technology Company \\
Beijing, China \\
zhenzhen\_wang@tjweather.com}
\and
\IEEEauthorblockN{Huiyuan Li}
\IEEEauthorblockA{Institute of Software, \\
Chinese Academy of Sciences \\
Beijing, China \\
huiyuan@iscas.ac.cn}
\thanks{\IEEEauthorrefmark{1}Corresponding authors are Yuan Liang (yuan\_liang@tjweather.com) and Xi Chen (chenxi@lasg.iap.ac.cn).}
}

\maketitle

\begin{abstract}
We present Super Dynamics on the Cube (SD3), the first operationally-proven, fully heterogeneous-accelerated global kilometer-scale model. Through a full-stack reconstruction in native CUDA C, SD3 introduces an ``Inline Physics'' architecture that deeply couples its non-hydrostatic cubed-sphere dynamical core with fast physical processes. This design maintains high physical consistency, ensures energy and mass conservation, and mitigates ``Grey Zone'' mismatches. Five years of continuous operational forecasting, together with validation against diverse observations, demonstrates the scientific fidelity of SD3. Recently, SD3 achieved an unprecedented 205.7 simulated days per day at global 1.5~km resolution using 24576 GPUs. SD3 establishes a hardware-agnostic, portable, and reproducible paradigm across diverse Exascale ecosystems. Beyond weather and climate research, SD3 offers a transformative paradigm for the HPC community and the next-generation Digital Twin Earth to establish a robust, high-performance blueprint for the future of Earth system modeling.
\end{abstract}


\vspace{1em}

\section{Overview}
Climate change has become a critical driver affecting the Earth system and human society. According to the Intergovernmental Panel on Climate Change (IPCC) Sixth Assessment Report (AR6), global warming is manifested not only in rising mean temperatures but also in the significantly increased frequency and intensity of extreme weather events \cite{ipcc}. Reliable climate simulations, especially precise projections of high-impact events like extreme precipitation and tropical cyclones (TCs), are paramount for global risk management and safeguarding lives. \par

However, current Global Climate Models (GCMs) face formidable challenges in simulating precipitation and associated extremes; in tropical regions, precipitation biases often overshadow the signals of climate projections \cite{palmer_scientific_2019}. To achieve accurate risk assessment and attribution of these meso-to-microscale physical processes, GCMs must transcend traditional resolution limits (typically 10--100~km) and shift toward Convection-Permitting Models (CPMs) with horizontal grid spacing below 4~km \cite{prein_review_2015}. At this scale, the traditional hydrostatic balance assumption becomes invalid as it fails to capture the intense vertical accelerations inherent in deep convective systems \cite{dudhia_nonhydrostatic_1993,kato_hydrostatic_1995}. Therefore, models must transition to more complex non-hydrostatic dynamical cores \cite{chen_control-volume_2013}. This shift introduces high-frequency sound waves that travel at extreme speeds, imposing much more stringent Courant-Friedrichs-Lewy (CFL) stability conditions on numerical schemes. This implies that as resolution increases, the total computational cost grows at a power-law rate \cite{xue2007numerical}, rapidly exceeding the capacity of existing computing systems and creating a severe bottleneck for high-resolution climate modeling. \par

Despite the dawn of the Exascale era, a profound ``structural mismatch'' persists between traditional modeling paradigms and emerging heterogeneous architectures. Massive legacy codebases, deeply rooted in Fortran and designed for Central Processing Unit (CPU)-centric processing, possess data layouts and memory access patterns that are fundamentally ill-suited for the massive parallelism of Graphics Processing Units (GPUs). Numerical modeling is currently undergoing a paradigm shift from traditional multi-core CPUs to GPU-based heterogeneous architectures \cite{klocke_computing_2025,fuhrer_near-global_2018,giorgetta_icon-model_2022,10.1145/3581784.3627044,zhang_lmarspy_2025}. However, this transition is not a simple code porting exercise; it represents a profound transformation touching the very logic of underlying design. \par

Current efforts to adapt these legacy codes to GPUs largely rely on directive-based approaches such as OpenACC or portable programming frameworks \cite{carter_edwards_kokkos:_2014,beckingsale_raja:_2019,medina_occa:_2014,ben-nun_stateful_2019}. While these methods offer a quick path to porting, they fail to redesign the underlying algorithms for modern hardware. As a result, these models often become ``memory-bound'' (starved for data) rather than ``compute-bound'', leaving the vast computational power of GPUs severely underutilized \cite{schulthess_reflecting_2019,govett_exascale_2024}. Furthermore, traditional models rely heavily on Double-Precision (FP64) calculations to maintain stability and physical consistency, whereas heterogeneous architectures offer significantly higher throughput for Single-Precision (FP32) or half-precision \cite{higham_new_2019,haidar_harnessing_2018}. While reducing precision can accelerate computation, it poses substantial risks in complex fluid dynamics, including numerical non-convergence, floating-point overflows, and grid-scale oscillations \cite{vana_single_2017,cotronei_single-precision_2020}. \par

In summary, the core problem is a dual challenge: climate science urgently demands kilometer-scale, non-hydrostatic simulations, but legacy software architectures cannot efficiently harness modern heterogeneous supercomputers to deliver them. To address this global challenge, Super Dynamics on the cube (SD3) was developed and demonstrated here. SD3 is derived from the ``Finite-Volume dynamical core on the Cubed-Sphere'' (FV3), a framework that has demonstrated exceptional TC forecasting performance in the HiRAM and HiFLOR models \cite{zhao_retrospective_2010,murakami_seasonal_2016}. Leveraging on this well-established foundation, SD3 implements a ground-up, full-stack reconstruction from Fortran to native Compute Unified Device Architecture (CUDA) C, specifically engineered for heterogeneous architectures. By introducing an innovative inline physics design, restructuring data organization, and implementing a mixed-precision strategy, SD3 overcomes memory bandwidth limitations and achieves unprecedented operational throughput for kilometer-scale climate modeling. \par

\vspace{1em}

\section{Current State of the Art}
In recent years, global climate modeling has undergone a paradigm shift from parameterized convection (10--100~km) toward explicitly resolving deep convection at kilometer-scale resolutions (1--5~km). International initiatives like DYAMOND \cite{stevens_dyamond:_2019,takasuka_protocol_2024} and nextGEMS \cite{segura_nextgems:_2025} have pioneered this transition, moving from short-term case studies to long-term, high-resolution climate integrations. \par

On the high-performance computing (HPC) front, two representative models define the current State-of-the-Art (SOTA) for kilometer-scale simulations: SCREAM and ICON. SCREAM, the Gordon Bell Prize winner for Climate Modelling in 2023, achieved global simulations at 3.25~km resolution on 32768 AMD GPUs using a Kokkos-based C++ refactoring \cite{10.1145/3581784.3627044}. More recently, ICON pushed the boundary further on the NVIDIA GH200-based JUPITER system, performing fully coupled global simulations at 1.25~km resolution and achieving a peak throughput of 145.7 simulated days per day (SDPD) \cite{klocke_computing_2025}. These milestones brilliantly demonstrate the technical feasibility of kilometer-scale climate modeling under extreme computing conditions. \par

However, these SOTA efforts primarily focus on demonstrating peak performance within narrow time windows, typically involving short-duration, ``hero-run'' experiments. While essential for assessing hardware limits, they provide limited insight into the long-term numerical stability and sustained operational efficiency required for realistic climate projections. The true challenge for kilometer-scale models lies in maintaining computational efficiency over long integrations, multiple initialization cycles, and complex physical feedback processes without crashing. Evaluating a model solely on peak throughput or a single-run scale is insufficient to characterize its practical scientific value. \par

To comprehensively measure both simulation scale and sustained robustness, we introduce the metric of Accumulated Degrees of Freedom (DOF·hr). This metric integrates horizontal grid density, vertical stratification, and the number of prognostic variables over the total successful integration time, formulated as: $$\text{DOF}\cdot\text{hr} = T_\text{total} \times \left( N_\text{grid} \times N_\text{lev} \times N_\text{var} \right) \text{,}$$ where $T_\text{total}$ is the total integration time in hours, $N_\text{grid}$ is the horizontal grid points, $N_\text{lev}$ is the vertical levels, and $N_\text{var}$ represents the number of prognostic variables. By focusing on the cumulative workload, DOF·hr serves as a rigorous proxy for numerical stability and software maturity, effectively distinguishing robust operational systems from experimental ``hero-runs''. \par

\begin{figure}[!hbt]
\centering
\includegraphics[width=\linewidth]{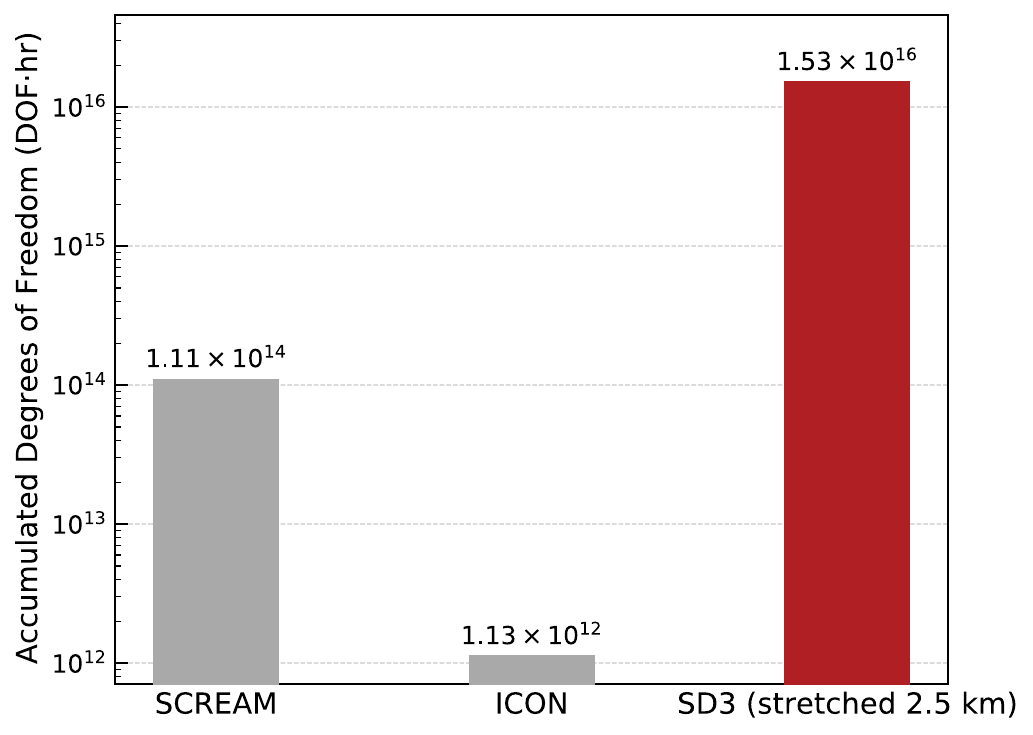}
\vspace{-2em}
\caption{Comparison of accumulated degrees of freedom (DOF·hr) for SCREAM, ICON, and SD3 (stretched 2.5~km) at convection-permitting resolution on heterogeneous supercomputing platforms.}
\label{fig:dof}
\end{figure}

Under this rigorous accounting, the limitations of current SOTA benchmarks become evident (Figure~\ref{fig:dof}). ICON's 1.25~km simulation, while achieving remarkable horizontal density ($3.36 \times 10^{8}$ cells per level), was conducted over a 3-hour window. With 90 vertical levels and 12.5 prognostic variables, its cumulative workload is approximately $1.13 \times 10^{12}$ DOF·hr. Similarly, SCREAM's 3.25~km quasi-climate experiment, integrated for 40~days (960~hours) with a grid configuration of $1024^2 \times 6$ points, 128 levels, adopts a spectral-element discretization that results in approximately 9 degrees of freedom per element per level. Accounting for 16 prognostic variables, this yields a total workload of $1.11 \times 10^{14}$ DOF·hr. While these figures represent the current apex of experimental capability, they highlight the episodic and short-lived nature of most kilometer-scale endeavors. \par

In stark contrast, SD3 represents a fundamental transition toward operational endurance at scale. Since 2021, SD3 has sustained an uninterrupted C1536\_r25 (2.5~km stretched over China and 6~km globally on average; 95 levels) production schedule of two initialization cycles per day, with each cycle executing 240-hour forecasts. Operating at convection-permitting resolution featuring $1536^2 \times 6 \times 95$ grid points and 13 prognostic variables, this continuous 1825-day deployment has accumulated a staggering $1.53 \times 10^{16}$ DOF·hr. By exceeding the experimental SOTA by two to four orders of magnitude (Figure~\ref{fig:dof}), SD3 demonstrates that it has not only conquered the exascale performance barrier but has also mastered the long-term numerical stability essential for reliable global climate and weather research. \par

\begin{table*}[!hbtp]
\caption{Performance comparison of SCREAM, ICON, and SD3 at kilometer-scale resolution on heterogeneous supercomputing platforms.}
\centering
\renewcommand{\arraystretch}{1.5}
\renewcommand{\tabularxcolumn}[1]{m{#1}}
\newcolumntype{Y}{>{\centering\arraybackslash}X}
\newcolumntype{s}{>{\hsize=.7\hsize}Y}
\begin{tabularx}{0.9\linewidth}{sssYYYYY}
\hline
\textbf{Model} & \textbf{Vertical Levels} & \textbf{Resolution (km)} & \textbf{Peak Throughput (SDPD)} & \textbf{Hardware} & \textbf{FP64 / FP32 Performance (TFLOPs)} & \textbf{HBM Bandwidth (TB/s)} & \textbf{CPU-to-GPU Bandwidth (GB/s)} \\
\hline
{SCREAM (2023)} & {128} & {3.25} & {458.2} & {32768 AMD MI250X on Frontier} & {47.9 / 47.9} & {3.2} & {72} \\
{ICON (2025)} & {90} & {1.25} & {145.7} & {20480 NVIDIA GH200 on JUPITER} & {34 / 67} & {4} & {900} \\
{SD3 (2026)} & {91} & {1.5} & {205.7} & {24576 GPUs on CNIS} & {32.7 / 65.5} & {1.8} & {64} \\
\hline
\end{tabularx}
\label{tab:perf}
\end{table*}

Furthermore, SD3 does not sacrifice peak speed for stability. In our latest full-scale deployment on 24576 GPUs on a China's New-generation Intelligent Supercomputer (hereinafter CNIS), SD3 achieved an unprecedented 205.7~SDPD at a global 1.5~km resolution, surpassing ICON's 145.7~SDPD (Table \ref{tab:perf}). This dual supremacy, setting a new high watermark in both peak throughput and sustained operational workload, demonstrates that SD3 has successfully bridged the critical gap from experimental feasibility to robust, practical applicability in Exascale climate modeling. \par

\vspace{1em}

\section{Innovations Realized}
\subsection{Super Dynamics on the Cube}
The non-hydrostatic dynamical core in SD3 is developed by TianJi Weather Science and Technology Company, engineered specifically for next-generation heterogeneous computing architectures. While inheriting the cubed-sphere grid discretization and vertical Lagrangian semi-implicit solver from the FV3 core \cite{lin_class_1994,lin_multidimensional_1996,lin_finitevolume_1997,lin_explicit_1997,lin_vertically_2004,putman_finite-volume_2007}, SD3 has undergone a deep reconstruction to meet the operational demands of kilometer-scale convection-permitting resolutions. \par

A pivotal innovation is the ``Inline Physics'' architecture, which facilitates the efficient coupling of fast and slow physical processes with the dynamical core. By embedding ``fast'' physical processes including 3D scale-aware moist convection, cloud microphysics, sub-grid orographic gravity wave, and turbulent drag directly into the dynamical core, SD3 effectively mitigates the resolution mismatch inherent in traditional parameterizations within the ``Grey Zone'' \cite{gerard_integrated_2007}. SD3 also accounts for flow-blocking induced by hundred-meter-scale sub-grid orography, enabling it to resolve complex topography and enhance the forecast skill in those regions. Furthermore, by separating the slow processes from the fast physics, the model allows for extended physical time steps via time-splitting techniques, significantly enhancing overall time-integration efficiency \cite{chen_efficient_2025}. The architecture adheres to conservation laws, ensuring the conservation of Total Moist Energy, momentum, and mass. \par

\subsection{Cubed-Sphere Grid}
SD3 employs an equi-distance cubed-sphere grid \cite{chen_lmars_2021,sadourny_conservative_1972}, which fundamentally resolves the pole singularity constraints inherent in traditional latitude-longitude grids caused by meridian convergence. By eliminating numerical singularities at the poles, SD3 bypasses the need for complex polar filtering, thereby removing the stringent CFL condition limits on time steps typically imposed by excessively small grid cells in polar regions. Furthermore, the high topological symmetry across the six tiles of the cubed-sphere grid ensures an exceptionally uniform distribution of computational load during parallel domain decomposition, which significantly enhances the scalability of large-scale parallel tasks. SD3 is also able to allocate more computational resources to localized weather systems by stretching the grid centered on target regions through the Schmidt transform \cite{schmidt_variable_1977}, demonstrating flexible multi-scale modeling capabilities. \par

\subsection{CUDA C Reconstruction}
To fully exploit the performance potential of heterogeneous computing architectures and acheive the highest efficiency possible, SD3 has undergone a full-stack deep reconstruction of its core computational components from Fortran to CUDA C. This comprehensive refactoring encompasses the dynamical core, physical schemes (including YSU planetary boundary layer \cite{hong_new_2010,wilson_modeling_2018} and RRTM radiation \cite{clough_atmospheric_2005}), Noah land surface model \cite{ek_implementation_2003}, and Mixed-Layer Ocean (MLO) model \cite{pollard_deepening_1973}, complemented by rigorous kernel-level operator optimization and performance tuning tailored for heterogeneous accelerators. \par

A primary challenge in Exascale computing is maintaining hardware portability without escalating code complexity. Unlike traditional models which hardcode platform-specific directives into computational kernels, SD3 employs a unified hardware-agnostic abstraction. All API differences between NVIDIA CUDA and AMD ROCm/HIP are encapsulated in a single compile-time macro header, mapping both backends to a consistent \texttt{\footnotesize{device*}} namespace (e.g., \texttt{\footnotesize{deviceMalloc}}, \texttt{\footnotesize{deviceError\_t}}). The entire scientific codebase is written in standard CUDA C and interacts exclusively with these abstract interfaces. Consequently, switching the compilation target from NVIDIA to AMD requires only a single build flag \texttt{\footnotesize{-DUSE\_CUDA}} (as shown in Figure~\ref{fig:code1}), ensuring that SD3's algorithms are truly portable and reproducible across diverse supercomputing ecosystems. \par

\begin{figure}[!hbt]
\centering
\begin{minted}[frame=lines,fontsize=\scriptsize]{cuda}
/* Hardware-Agnostic Abstraction */

#ifdef USE_CUDA
  #include <cuda_runtime.h>
  #define deviceSuccess cudaSuccess
  #define deviceGetErrorString cudaGetErrorString
  #define deviceMalloc cudaMalloc
  #define deviceMallocHost cudaMallocHost
  #define deviceError_t cudaError_t
  #define deviceHostFree cudaFreeHost
#else
  #include <hip/hip_runtime.h>
  #define deviceSuccess hipSuccess
  #define deviceGetErrorString hipGetErrorString
  #define deviceMalloc hipMalloc
  #define deviceMallocHost hipHostMalloc
  #define deviceError_t hipError_t
  #define deviceHostFree hipHostFree
#endif

void *deviceAllocate(size_t len, const char *str)
{
  void *ptr = NULL;
  deviceError_t err = deviceMalloc(&ptr, len);
  if (err != deviceSuccess)
  {
    printf("%s device malloc error!\n", str);
    fflush(stdout);
    printf("error string=%s\n",
           deviceGetErrorString(err));
  }
  assert(ptr != NULL);
  return ptr;
};
\end{minted}
\vspace{-1em}
\caption{Illustration of the hardware-agnostic abstraction.}
\label{fig:code1}
\end{figure}

The reconstruction includes several low-level optimizations tailored for heterogeneous accelerators: \par

\begin{itemize}
\item Low-level Instruction Optimization and Register Management: Addressing the characteristics of Vector General-Purpose Registers (VGPRs), which offer ultra-fast access but limited capacity and lack native cross-thread visibility, SD3 moves beyond the conventional approach of using ``redundant calculation'' to mimick data sharing by employing inline assembly to encapsulate low-level hardware instructions. This enables efficient cross-lane communication, minimizing computational redundancy while fully tapping into the advanced instruction set's throughput potential.

\item On-chip Memory Pressure Optimization: By leveraging macro definitions to hardcode algorithmic constants as immediate values within the instruction stream, SD3 significantly reduces the footprint of constants on registers and constant memory, thereby alleviating resource pressure on the computing devices.

\item 3D Addressing Abstraction: An index mapping layer was constructed via macros to maintain the high performance of 1D flat memory access while enabling intuitive 3D logical addressing (demonstrated in Figure~\ref{fig:code2}). This strategy balances hardware execution efficiency with the code maintainability of complex physical algorithms.

\begin{minipage}{\linewidth}
\vspace{1.5em}
\centering
\begin{minted}[frame=lines,fontsize=\scriptsize]{cuda}
/* 3D Addressing Abstraction */

#define gz(i, j, k)                           \
  gz[(k - 1) * nxd * nyd + (j - jsd) * nxd +  \
     (i - isd)]
#define dz2(i, j, k)                          \
  dz2[(k - 1) * nxc * nyc + (j - js) * nxc +  \
      (i - is)]

int ij = blockIdx.x * blockDim.x + threadIdx.x;
int k = blockIdx.y + 1;
int i = ij % nxc + is;
int j = ij / nxc + js;

// core computation
if ((j >= js && j <= je)
    && (i >= is && i <= ie)
    && (k >= 1 && k <= km))
{
  dz2(i, j, k) = gz(i, j, k + 1) - gz(i, j, k);
}
\end{minted}
\vspace{-1em}
\captionof{figure}{Illustration of the 3D addressing abstraction.}
\label{fig:code2}
\vspace{1em}
\end{minipage}
\vfill

\item Unified Memory Pool and Fragmentation Control: Frequent dynamic memory allocations and deallocations on the device are computationally expensive and trigger memory fragmentation, which often leads to allocation failures during long-term operational runs. SD3 addresses this by implementing a global pre-allocated memory pool on the device. Initialized during model startup, this large contiguous array serves all temporary variables in the model. Variables access the pool through typed pointer offsets, supporting multiple data types while maintaining zero runtime allocation overhead.
\end{itemize}

The memory pool strategy is particularly critical for the RRTM radiation scheme, which presents the most significant memory bottleneck at kilometer-scale resolutions. The high-dimensional temporary arrays required for RRTM would exceed available GPU memory if processed concurrently. SD3 resolves this by implementing a tiled radiation scheduling strategy: the horizontal grid is partitioned into tunable ``blocks'', and the pre-allocated memory pool is sized to accommodate the requirements of exactly one block. By reusing these buffers across successive block iterations, SD3 balances memory footprint against computational throughput, allowing the model to adapt to various GPU memory capacities while maintaining high-throughput execution. \par

As a result of these optimizations, the 3D computational kernels in SD3 achieved an average speed up of 16.15\%, with peak performance gains reaching 26.05\%. Recognizing that single-precision (FP32) throughput significantly outpaces double-precision (FP64) on modern accelerators, SD3 adopts a mixed-precision strategy and achieves a substantial boost in throughput by optimizing most compute-intensive variables from float64 to float32 while maintaining physical consistency. The initialization and I/O modules retain the original Fortran framework to ensure compatibility. \par

\subsection{Asynchronous I/O and Communication}
To address I/O bottlenecks in ultra-large-scale simulations, SD3 utilizes a parallelized scheme based on ``I/O Aggregators''. This approach avoids the pitfalls of centralized ``N-to-1'' I/O (which causes bandwidth saturation) and individual ``N-to-N'' output (which generates excessive small files and stresses performance). Instead, it employs a Process Group-based aggregated output mode. By aggregating data within groups for root-process writes, SD3 significantly reduces wall-clock time penalties during high-resolution operational runs. \par

Driven by its operational demands, SD3 prioritizes end-to-end timeliness through a streamlined data lifecycle where each day of simulation output is processed immediately for user delivery. This high-throughput workflow is enabled by a native, distributed post-processing toolchain that performs real-time interpolation directly from scattered sub-tile files. Bypassing the traditionally I/O-intensive and time-consuming ``combine'' phase, SD3 eliminates a critical bottleneck, ensuring a seamless transition from raw simulation to actionable diagnostic analysis. \par

SD3 employs GPUDirect RDMA (GDR)-enabled MPI communication, enabling direct inter-device data transfers across compute nodes without host-memory staging, thereby reducing data-movement costs and improving parallel communication efficiency. In addition, SD3 implements carefully designed asynchronous computing and communication masking strategies to maximize heterogeneous node utilization. Through asynchronous stream design, MPI collective communications overlap with accelerator-side computational kernels, effectively hiding communication latency. \par

\subsection{Physics-AI Synergy: Enhancing Medium-Range Reliability}
\begin{figure}[!hb]
\centering
\includegraphics[width=\linewidth]{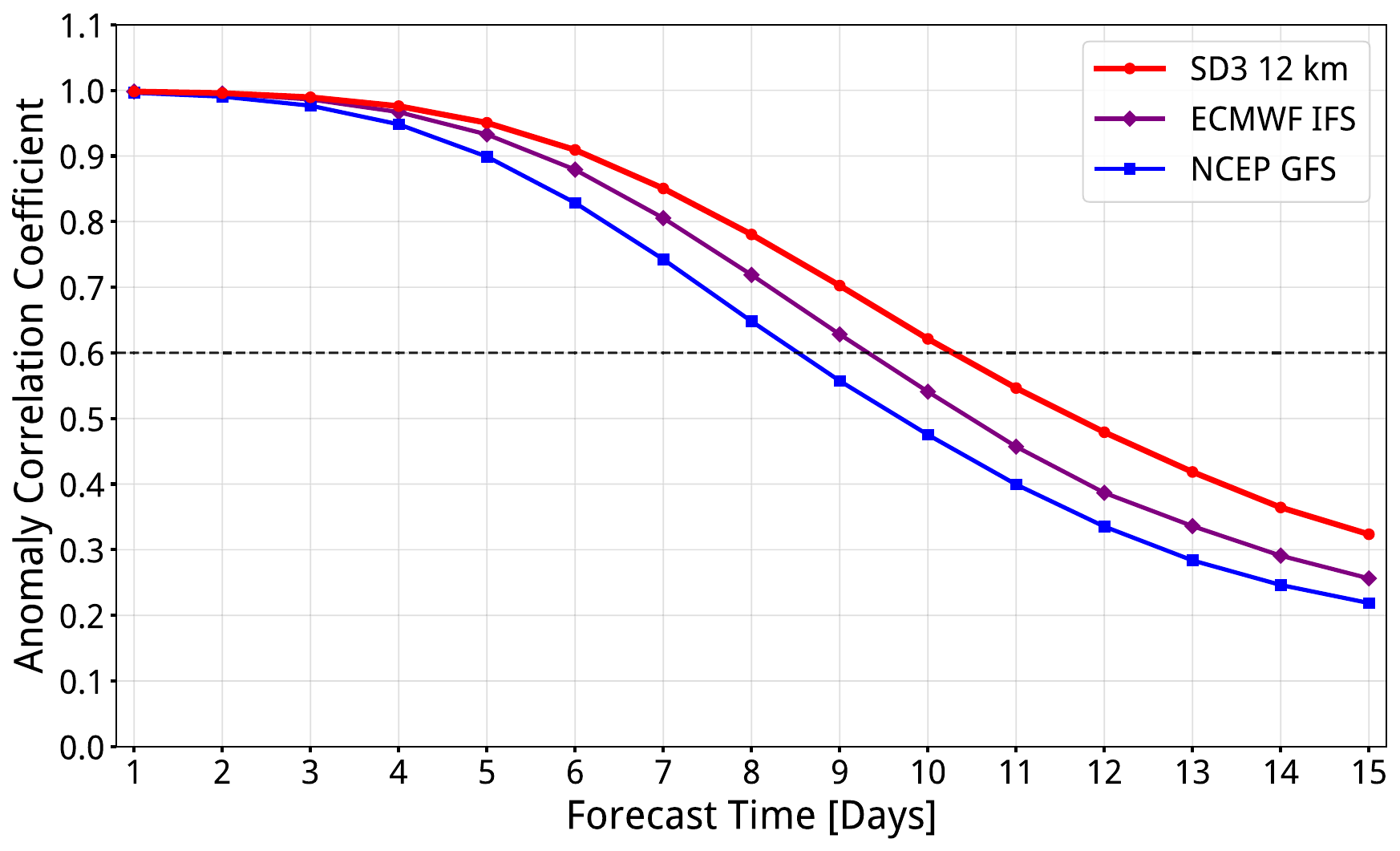}
\vspace{-2em}
\caption{Comparison of SD3 12~km, ECMWF IFS, and NCEP GFS in terms of Northern Hemisphere H500 ACC as a function of forecast lead time. Verified against ECMWF ERA5 reanalysis (March 2025--February 2026), the evaluation includes 730 cases from both 00Z and 12Z cycles. SD3 12~km consistently outperforms both operational models across all lead times.}
\label{fig:acc}
\end{figure}

For operational forecasts, SD3 employs a Hybrid Physics-Artificial Intelligence (AI) Integration scheme to redefine medium-range reliability at high resolutions. Addressing the rapid skill degradation of large-scale atmospheric patterns in traditional numerical weather model, SD3 innovatively fuses heterogeneous ensembles of global 12~km physical simulations with AI model outputs. By utilizing Machine Learning for dynamic weighting of each member, SD3 constructs high-quality large-scale nudging fields to drive its operational runs (including global 12~km run). This synergy combines the structural advantages of ensemble-based circulation with the high-fidelity resolving power of the physical model. As shown in Figure~\ref{fig:acc}, SD3 has achieved a breakthrough in maintaining a Northern Hemisphere 500~hPa geopotential height anomaly correlation coefficient (H500 ACC) above 0.6 at Day 10 (730 cases, March 2025--Feb 2026), reaching a world-leading global forecast skill. \par

\section{How Performance Was Measured}
\subsection{Model Configuration}
Two global cubed-sphere grid configurations are employed for the scalability experiments: a high-resolution 1.5~km (C6144) grid and its low-resolution 12~km (C768) counterpart, both featuring 91 vertical layers. The integration utilizes a split-explicit time-stepping scheme with a time step of 90~s and 225~s for the 1.5~km and 12~km grids, respectively. Within the dynamical core, the vertical Lagrangian remapping and tracer update intervals are set to 22.5~s and 2.25~s for the 1.5~km grid, as well as 112.5~s and 18.75~s for the 12~km grid. Atmospheric initial conditions are derived from the ECMWF IFS analysis, while land surface states are initialized using the NCEP GFS analysis. The model incorporates a MLO component, where the simulated sea surface temperatures (SST) are relaxed towards climatological values following \cite{harris_gfdl_2020}. \par

\subsection{System Details}
All experiments are conducted on CNIS, where each compute node comprises two 64-bit CISC-based server processors (64 cores per processor, 2.4~GHz), eight SIMT-based GPU computing accelerators, and 512~GB system memory. Each accelerator features 64~GB of High Bandwidth Memory (HBM) with a peak bandwidth of 1.8~TB/s. Intra-node connectivity is through PCIe Gen5, while inter-node communication relies on a 1.6~Tb/s proprietary InfiniBand-like native Remote Direct Memory Access (RDMA) interconnect. The system operates on Anolis OS 8.9. The software environment includes a GPU programming suite compatible with mainstream GPGPU API standards. Inter-accelerator communication utilizes a collective communications library compatible with RCCL APIs. \par

\subsection{Performance Metrics}
Performance is quantified using SDPD, which describes the model throughput in units of simulated time versus actual wall-clock time. We derive the SDPD from the time it takes to simulate 24~hours in the 1.5~km configuration, excluding the initialization time. Initialization requires a measurable amount of time during our short benchmark runs; however, in a production environment where the model runs continuously for days or months, its contribution to the overall runtime is relatively small, which is why we exclude it here. \par

The simulation time includes the super dynamics (dynamical core and inline fast physics) and all physical parameterizations (slow atmospheric physics, land, and mixed layer ocean processes). Timings are captured using CPU timestamp counters. Measurements are taken across all MPI ranks, and the maximum value is presented to account for any potential load imbalance. \par

\begin{figure}[!b]
\centering
\vspace{-1em}
\includegraphics[width=\linewidth]{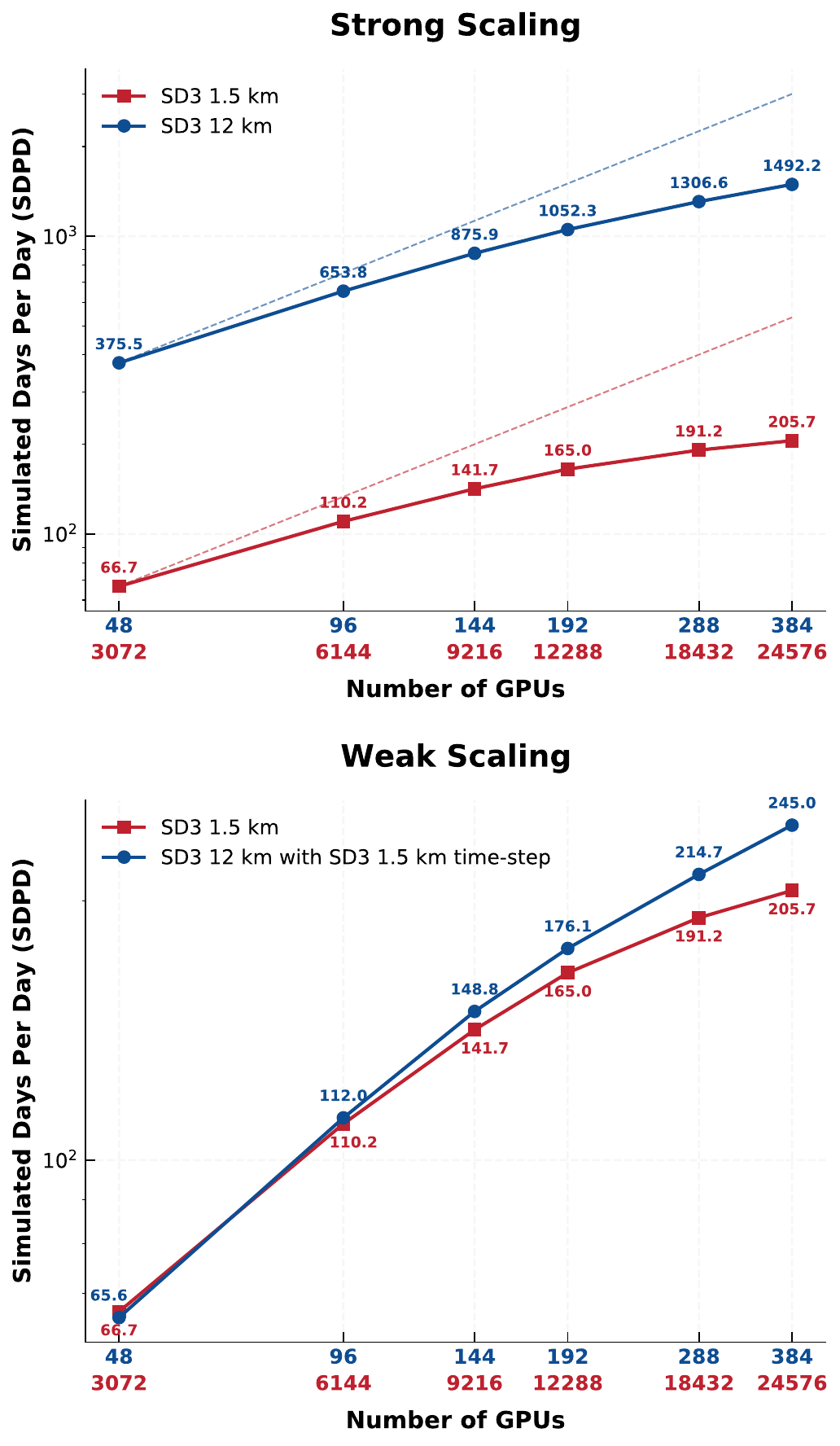}
\vspace{-2em}
\caption{Strong (top) and weak (bottom) scaling performance of SD3 at 12~km and 1.5~km resolutions on the target GPU platform, with throughput measured in SDPD as a function of GPU counts.}
\label{fig:perf}
\end{figure}

\vspace{1em}

\section{Performance Results}
On CNIS, each node is equipped with 8 GPUs; therefore, we configured each node with 8 MPI ranks and 16 threads. Strong scaling results were evaluated at both 1.5~km and 12~km resolutions, demonstrating relatively high parallel efficiency, as shown in Figure~\ref{fig:perf} (top). For 12~km runs, using 48 to 384 GPUs yielded a SDPD range of 375.5 to 1492.2 with a parallel efficiency of 49.7\%. For 1.5~km runs, scaling from 3072 to 24576 GPUs achieved a SDPD of 66.7 to 205.7 with an efficiency of roughly 38.5\%. Strong scaling performance is primarily limited by the increased relative cost of communication as the computational workload per GPU decreases. \par

Weak scaling performance was characterized by comparing the 1.5~km and 12~km configurations, with both experiments employing the 1.5~km time step, as illustrated in Figure~\ref{fig:perf} (bottom). Since the 1.5~km resolution represents an 8-fold increase in horizontal resolution over 12~km, the total number of grid points increases by a factor of 64. To maintain a constant computational load per device, the GPU count for 1.5~km runs was set to 64 times that of the 12~km configuration. The results show excellent weak scaling behavior: at the 12~km resolution using 384 GPUs, the SDPD was 245.0, while at the 1.5~km resolution using 24576 GPUs, the SDPD reached 205.7, resulting in a weak scaling efficiency of 83.9\%. Weak scaling is similarly affected by the rising communication-to-computation ratio as the total number of GPUs increases. \par

By adopting GDR-enabled MPI, SD3 eliminates device-host memory-copy overhead during communication. In the SD3 1.5~km experiment conducted on 24576 GPUs, this optimized implementation improved SDPD by 19\% compared with the conventional CPU-based MPI version. At the 1.5~km resolution, the restart file size was approximately 1.5~TB. Utilizing 24576 GPUs, the write speed of restart files reached approximately 200~GB/s. \par

\begin{figure*}[!t]
\centering
\includegraphics[width=0.88\linewidth]{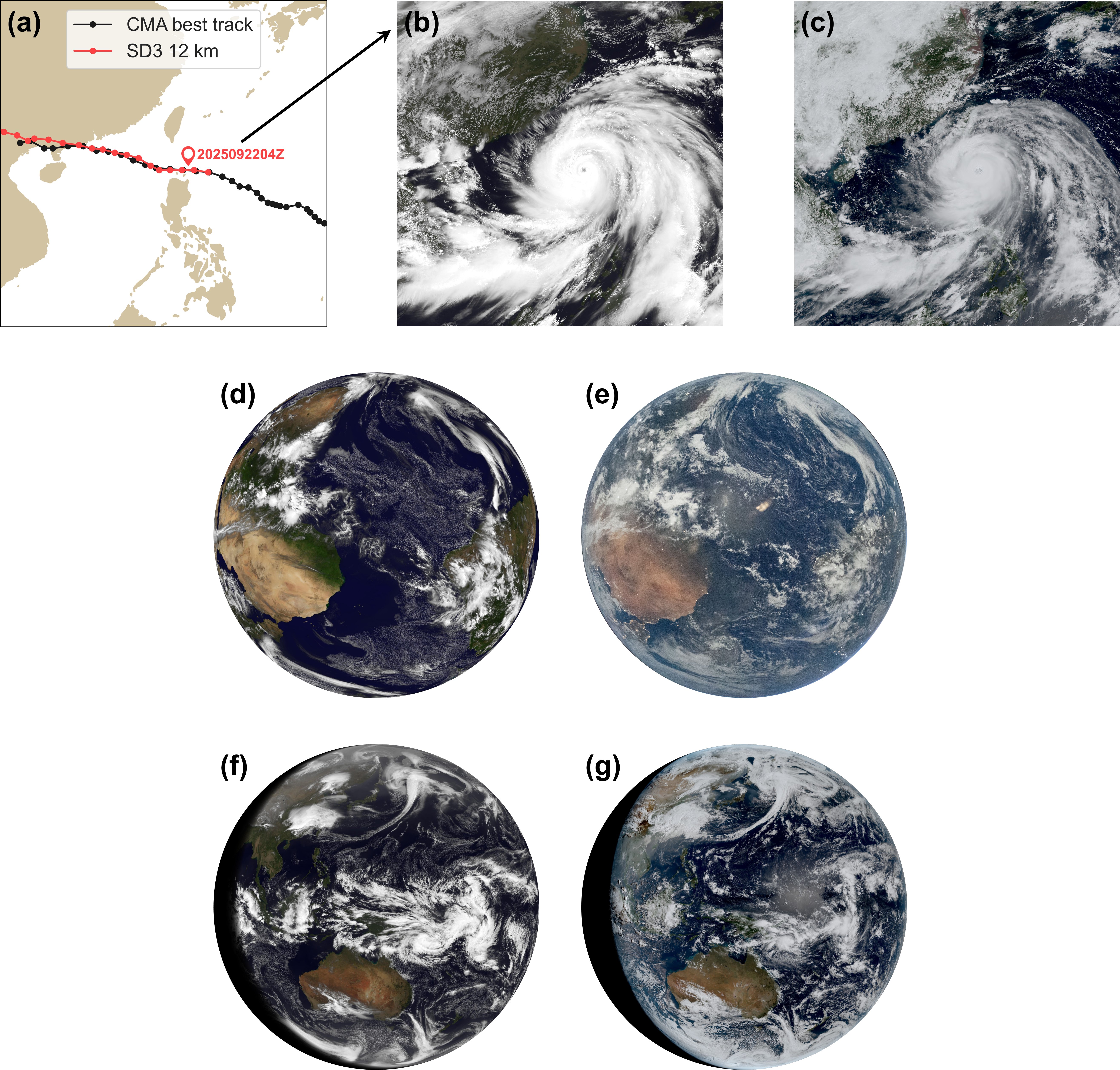}
\caption{Multi-scale validation of SD3 against observations for TC simulation and global cloud structures. (a) Super Typhoon Ragasa track forecast from the SD3 12~km operational forecast initialized at 2025092112Z, compared with the China Meteorological Administration (CMA) best-track. (b--c) Snapshot of Ragasa at 2025092204Z: (b) SD3 simulation at 1.5~km resolution initialized at 2025092112Z; (c) Himawari-9 true color imagery. (d--g) Global view at 2026040300Z during NASA Artemis II mission: (d) Predicted cloud fields by SD3 1.5~km simulation initialized at 2026040200Z, mimicking (e) the photograph captured by astronaut Reid Wiseman; (f) Same SD3 1.5~km simulation as viewed from the Himawari-9 perspective; (g) Himawari-9 true color imagery.}
\label{fig:cloud}
\vspace{-1em}
\end{figure*}

\vspace{1em}

\section{Implications}
In the era of kilometer-scale seamless forecasting, the transition toward a unified dynamics-physics architecture represents a significant scientific milestone. At convective-permitting scales, traditional methods where dynamics and physics are integrated sequentially often introduce inconsistent numerical feedbacks and risk violating the conservation of energy and moisture. SD3 addresses this by deeply coupling the dynamical core with fast physical processes within a unified temporal framework. This approach minimizes numerical splitting errors and allows for a more consistent representation of non-linear atmospheric interactions, fundamentally enhancing the fidelity of extreme weather simulations. By the full-stack CUDA C reconstruction for Exascale heterogeneous architectures, SD3 achieves a transformative leap in parallel efficiency. Its global variable-resolution design further enables the precision-targeted allocation of computational resources, making ultra-high-resolution local simulations computationally feasible without sacrificing the global context. \par

Beyond climate science, SD3 offers profound implications for the broader HPC community, directly addressing the critical challenge of generalizability. By embedding fast physical processes directly into the dynamical core, SD3 maximizes data reuse within GPU registers and shared memory, dramatically reducing costly global memory accesses. Furthermore, while demonstrated on a specific heterogeneous cluster, SD3's codebase is fundamentally hardware-agnostic. Its native CUDA C implementation, explicit register management, and asynchronous I/O strategies can be seamlessly ported to other mainstream Exascale platforms (e.g., NVIDIA or AMD based systems), ensuring that these algorithmic innovations can benefit a wide range of Exascale applications. \par

From an operational perspective, SD3 has established itself as the first scientifically verifiable and robust tool for global kilometer-scale atmospheric modeling, proven by five years of continuous operational runs at convection-permitting scales over China (https://www.tjweather.com). While many international benchmarks (e.g., HighResMIP) remain at \textasciitilde 25~km resolutions, SD3 effectively bridges the convection-permitting gap, significantly enhancing the representation of meso-scale systems such as Category 4 and 5 TCs and mitigating the spatial blurring and intensity underestimations typical of coarser models \cite{chen_present_2024,roberts_projected_2020}. \par

The fidelity of SD3 simulations is demonstrated by real-world observations (Figure~\ref{fig:cloud}). In the case of Super Typhoon Ragasa in 2025, the SD3 12~km operational forecast initialized at 2025092112Z aligned closely with the CMA best-track (Figure~\ref{fig:cloud}a), while the 1.5~km simulation at the 16th forecast hour resolved a well-defined eye-wall structure and spiral bands in excellent agreement with Himawari-9 satellite imagery (Figure~\ref{fig:cloud}b--c). SD3's reliability is further corroborated during the recent NASA Artemis II mission. The 24th-hour forecast at 1.5~km resolution showed exceptional fidelity to the historic ``Hello, World'' full-Earth photograph captured by astronaut Reid Wiseman on April 3 (Figure~\ref{fig:cloud}d--e). Another view at the same time matched closely with Himawari-9 (Figure~\ref{fig:cloud}f--g). Together, these validations establish the reliability required to translate SD3's high-resolution forecasts into operational services, scientific research, and public-interest applications. \par

This operational maturity has generated tangible benefits for both industrial production and everyday life. SD3 forecast data are openly available through the AWS Open Data Program (\url{https://registry.opendata.aws/dynamical-tj-nwp/}). To date, SD3 has supported renewable-energy operations in 18 countries by providing high-resolution wind and solar forecasts that enable timely decision-making for market bidding, dispatch scheduling, and real-time grid balancing. SD3's kilometer-scale forecasts also facilitate transportation planning by identifying rainfall and storm-related hazards affecting roads, railways, and ports, as well as providing localized guidance for agricultural decisions concerning planting, irrigation, and harvesting. Moreover, SD3's fine vertical resolution captures the atmospheric structure needed for low-altitude economy, supporting flight-weather assessment, route-specific risk evaluation, and flight-planning decisions. \par

Building on its operational applications, SD3 serves as an open scientific resource for weather research and offers substantial potential for future climate studies. SD3 data have already supported a range of weather-related investigations, including analyses of extreme precipitation events \cite{2022029A,li_evaluation_2023} and the development of AI forecasting methods \cite{Q24077}. SD3 data have also contributed to interdisciplinary health studies linking temperature variability to epidemic risk. For future climate research, SD3 offers a promising paradigm for high-resolution climate modeling that transcends the limitations of traditional dynamical downscaling. While the integration of a fully 3D ocean model is currently underway, SD3 has already demonstrated its potential as a high-performance scientific tool in AMIP-style experiments, including DYAMOND3 \cite{takasuka_protocol_2024} and HighResMIP2 \cite{roberts_high-resolution_2025}. Unlike traditional one-way or two-way nested frameworks coupling global and regional models, SD3 achieves local refinement via variable-resolution technology within a unified global dynamical core. This avoids the lateral boundary inconsistencies inherent in nested configurations and mitigates the associated systematic errors \cite{inatsu_gcm_2012,vannitsem_one-way_2005}, enabling seamless coupling from planetary-scale circulation down to localized convective processes. This framework not only inherits the precision of dynamical downscaling but also overcomes the constraints of scale separation and boundary processing, providing a transformative technical pathway for high-resolution climate projections and the study of extreme events (e.g., \cite{lorenzo_global_2025,sieck_weather_2021,zhao_crucial_2024}). \par

Looking ahead, the scalable architecture of SD3 provides a clear pathway toward sub-kilometer global simulations. More precisely, 3~km is only deep convection resolving. To resolve shallow convection (cloud depth generally less than 3~km) the horizontal resolution would need to be on the order of 10--100~m \cite{matheou_fidelity_2011,schalkwijk_weather_2015}. Our ongoing work targets grid spacings below 1~km, which would enable the explicit resolution of individual convective cells and boundary-layer turbulence at the global scale, a frontier that has remained computationally inaccessible until now. By pushing the boundaries of both computational efficiency and physical fidelity, SD3 lays the groundwork for the next generation of Digital Twin Earth systems. \par

\vspace{1em}
\printbibliography

\end{document}